
\magnification=1200
\overfullrule=0pt
\baselineskip=20pt
\parskip=0pt
\def\dag{\dagger}
\def\del{\partial}

\def\g{\gamma}     
\def\d{\delta}     
\def\e{\epsilon}   
\def\z{\zeta}

\def\l{\lambda}    \def\L{\Lambda}
\def\m{\mu}	   
\def\n{\nu}

\def\p{\pi}        \def\P{\Pi}
\def\r{\rho}

\def\f{\phi}       
       
\def\y{\psi}       
\def\w{\omega}     
\def\br{\langle}
\def\ke{\rangle}
\def\ve{\vert}

\def\ba{\cal A}
\def\bb{b}
\def\adel{{\hbox{$\> \buildrel {\leftrightarrow}\over {\partial}\>$}}}

{\settabs 5 \columns
\+&&&&CCNY-HEP-93/1\cr}
\bigskip
\centerline{\bf THE ELECTROMAGNETIC INTERACTIONS OF ELECTRONS }
\centerline{\bf IN THE LOWEST LANDAU LEVEL}
\bigskip\bigskip
\centerline{Rashmi Ray$^1$ and B. Sakita$^2$}
\bigskip

\centerline{ Physics Department, City College of the City University of New
York}
\centerline{ New York, NY 10031}
\bigskip
\centerline{\bf Abstract}
\bigskip
Starting from a system of planar electrons in a strong magnetic field
normal to the plane, interacting with perturbing electromagnetic fields,
an effective Lagrangian for the fermions in the lowest Landau level (L.L.L.)
has been derived. By choosing a suitable background electrostatic potential,
an incompressible droplet of these electrons is constructed. The gauge
invariant
effective Lagrangian for the electrons in the L.L.L. is shown to split
naturally into a $1+1$ dimensional Lagrangian for the electrons on the
surface of the
droplet and into a $2+1$ dimensional gauge-field Lagrangian
representing the
contribution of the interior of the droplet. Upon bosonization,
the former represents the surface vibrations of the droplet.
Individually neither of these
two actions is gauge invariant, but it is shown that the gauge dependence
from the two pieces cancels out. This demonstrates that the edge degrees of
freedom are essential for maintaining gauge invariance.
\vfill
\noindent{$^1$E-mail address: rashmi@sci.ccny.cuny.edu.}

\noindent{$^2$E-mail address: sakita@sci.ccny.cuny.edu.}
\vfill\eject
\centerline{\bf I. Introduction}
\bigskip

The importance of the edge states for the observability of the Quantum
Hall effect was first pointed out by Halperin [1]. More recently, Wen [2],
Stone [3] and Fr\"ohlich and Kerler [4]
have delineated the role of these edge waves in maintaining
the electromagnetic $U(1)$ invariance of the system of planar electrons
confined to a droplet and placed in a strong magnetic field orthogonal
to the plane. They noted that when perturbative electromagnetic potentials
are coupled to the electrons in the droplet, the fermionic degrees of
freedom may be integrated out to obtain an effective action defined over
the domain of the droplet. This effective action generically contains
an Abelian Chern-Simons
term with its coefficients given by the quantized Hall conductance. Further it
is well known that a Chern-Simons term defined on a compact space is gauge
non-invariant, the non-invariance manifesting itself through a surface
term.

To restore gauge invariance, Wen [2] postulated a boundary action, expressed as
a bilinear in the perturbing potential, the coefficient of the bilinear
being the current-current correlator of the boundary current. If one requires
that the gauge-variation of this boundary action should cancel against that
of the Chern-Simons term, one gets conditions on the correlator.
This condition enables one to show that the low energy excitations of the
boundary Hamiltonian are massless. Further, the components of the currents in
the momentum space are the creation operators for these massless excitations
from the vacuum state.

The light-cone components of these currents satisfy the chiral $U(1)$
Kac-Moody algebra. Wen gives a field theory whose currents satisfy this
algebra. It is a theory of chiral fermions living on the boundary and
interacting with the perturbative electromagnetic fields on the boundary
in a fashion dictated by minimal coupling. But since chiral fermions in
$1+1$ dimensions are equivalent to chiral bosons, a bosonic realization
can also be given. Such a bosonic construction has been given by
Stone [3].

  Iso, Karabali and Sakita [5] have studied the   2+1 dimensional
non-relativistic
fermionic theory and have bosonized the theory from the onset. In the fermionic
theory, the strong
magnetic field projects the electrons to the lowest Landau level (L.L.L.). In
the bosonic language this L.L.L. condition leads to a classical configuration
in the form of a droplet whose shape is determined by the confining
potential. Namely, from $ A_0 (x,y) = \m $ we obtain a curve
$y=y(x)$ which is the boundary of the droplet.
 The bosonic fluctuations around this classical configuration
are precisely the chiral bosonic edge waves. The vibration of the surface just
means the excitation of electrons from
just below the Fermi surface to just above it. Since the energy levels here
are quasi-continuous, the excitations are gapless.
The magnetic field causes the electrons to rotate in a given direction and
thus the edge waves are chiral.
This present work can be taken to be an analysis of the same problem
in the presence of additional perturbative electromagnetic fields.

In previous works on the edge waves of the Quantum Hall droplet, the boundary
degree of freedom appears as a postulated construct designed to maintain gauge
invariance. Our purpose here is to demystify the origin of the edge waves. We
start from the original fermionic degrees of freedom and extract those that
describe the edge waves. On doing this we see that the bulk of the droplet is
naturally described by an effective Lagrangian involving the perturbing
potential. The electromagnetic interaction of the edge waves is also obtained
and we comment on the $U(1)$ gauge-invariance of this effective description.

Let us consider a system of planar (2+1 dimensional) electrons in a magnetic
field $B$ set up orthogonal to the plane. This magnetic field creates Landau
levels on the plane. Each Landau level has a degeneracy given by
$B/2\p$ per unit area. The gap between Landau levels is given by
$\w=B/m$ where $m$ is the mass of an electron.
Working with the single-particle Hamiltonian
$$h_0={1\over{2m}}(\vec p-\vec A)^2, \eqno(1.1)$$
where $\vec A=B(0,x)$ in the Landau gauge,
we see that the wave function for the nth Landau level
is given by
$$\br\vec x\ve
n,X\ke=\sqrt{B/2\p}(B/\p)^{1/4}{1\over\sqrt{2^nn!}}e^{iBXy}e^{-{B\over2}(x-X)^2}H_n(\sqrt B(x-X)),\eqno(1.2)$$
where $X$, the centre of the classical cyclotron motion
is arbitrary and is a measure of
the degeneracy. The corresponding eigen-energy is $E_n=(n+{1\over2})\w$ which
contains no reference to $X$. So given $n$, the index of the Landau level,
 the electron can be anywhere on the plane, depending on the value of $X$
chosen. However, if an uniform electric field is also turned on, say in the
$x$-direction, where $\vec E=(E,0)$, the energy eigenvalues are
$$E_n(X)=(n+{1\over2})\w+EX-E^2/2m\w^2.\eqno (1.3)$$
The corresponding eigenfunction is
$$\br\vec x\ve
n,X\ke=\sqrt{B/2\p}(B/\p)^{1/4}{1\over\sqrt{2^nn!}}e^{iBXy}e^{-{B\over2}(x-X-X_0)^2}H_n(\sqrt B(x-X-X_0)),\eqno(1.4)$$
where $X_0\equiv-E/m\w^2$.  So the degeneracy in $X$ is lifted and the
electron will drift to large negative values of $X$ to minimize the energy.
We can use this external electrostatic potential $A_0=Ex$ to define a
droplet of electrons. We fill all the single-particle states with $n=0$
and $X\leq0$. Physically, these electrons will exist in the negative
half-plane. We call this a droplet and identify the y-axis as the edge of
the droplet.

In a second-quantized description of this system, we choose as the ground
state, the state in the Fock-space with all the $X\leq0$ for $n=0$ filled.
So, the Fermi-surface in this case coincides with the physical edge of the
droplet. The droplet is incompressible as all the electronic states in it are
filled. The only way to excite it perturbatively  is to excite electrons with
$X$ small and negative to small and positive values of $X$. With reference to
the ground state defined earlier, this corresponds to producing neutral
electron-hole pairs. This leads to a deformation of the linear profile of
the edge . These are the edge excitations described in the literature.
Due to the magnetic field, they are chiral in nature, as has been mentioned
earlier.

In this work, we wish to extract the effective Lagrangian for these excitations
and the
nature of their electromagnetic interactions from the original
microscopic Lagrangian of
the planar fermions. Further, we shall investigate the role of the edge
excitations in preserving the original electromagnetic $U(1)$ gauge
invariance of the system. For this purpose we derive the effective Lagrangian
for the electrons in the L.L.L. , interacting with perturbative external
electromagnetic potentials, whose typical frequency is much less
than the cyclotron energy and the inverse of the magnetic length. Thus we keep
terms only upto
the quadratic in the electromagnetic potentials and consistently drop
the higher spatial and time derivatives of these potentials.
\bigskip
\centerline{\bf II.  Computing the effective action for the lowest Landau
level.}
\bigskip
The kinematics of electrons in the L.L.L. has already been discussed by
Girvin and Jach [6]. However, to make the discussion reasonably self-contained,
we introduce our own notation in the following.

The Lagrangian for planar electrons in a magnetic
field normal to the plane is
$$ \int d \vec x\ \y^\dag (\vec x, t) \biggl(i \del_t -h_0\biggr)
\y (\vec x, t). \eqno(2.1)$$
The corresponding action is given by
$$-S_0=\br\y \ve \hat{p}_t +h_0 \ve \y \ke, \eqno(2.2)$$
where $\br t \ve  \hat{p}_t = -i \del_t \br t \ve $
and $\y(\vec x,t)\equiv\br\vec x,t\ve\y\ke$ is the Schr\"odinger wave field for
electrons. The single-particle Hamiltonian, $h_0$, is
$$h_0={1\over{2m}}\vec \P^2.$$
$\vec \P \equiv \vec p-\vec A$, and $\vec A \equiv B (y,-x)$ is the vector
potential for the magnetic field in the symmetric gauge. $B$ is of
dimension $(\rm mass)^2$. So there are two parameters with mass dimension,
$\w={B/m}$ and $\sqrt B$, both of which are much greater than the typical
frequency of the external electromagnetic fields.  Define $\hat \p \equiv
{1\over\sqrt{2B}}(\hat \P^x-i\hat \P^y)$ and
$\hat \p^\dag \equiv {1\over\sqrt{2B}}(\hat \P^x+i\hat \P^y)$.
So, $\hat \p $ and $\hat \p^\dag $ are dimensionless. Their commutator is
$$ [\hat \p ,\hat \p^\dag ]=1$$
Dropping the zero-point energy, the Hamiltonian is written as
$$h_0=\w\hat \p^\dag \hat \p \eqno(2.3)$$
Further, we define the guiding centre coordinate operators
$$\hat X\equiv \hat x-{1\over B}\hat \P^y\qquad\hbox{and}\qquad\hat Y\equiv
\hat y+{1\over B}\hat \P^x$$
and their holomorphic form
$$\hat a\equiv \sqrt {B/2}(\hat X+i\hat Y)\qquad\hbox{and}\qquad\hat a^\dag
\equiv \sqrt {B/2}(\hat X-i\hat Y)$$
The commutators are $$[\hat X,\hat Y]={i\over B}\qquad\hbox{and}\qquad[\hat a,
\hat a^\dag ]=1\eqno(2.4)$$
$\hat a$ and $\hat a^\dag $ are dimensionless. Further,
$$[\hat a, \hat \p ]=[\hat a, \hat \p^\dag ]=[\hat a^\dag, \hat \p ]=[\hat
a^\dag, \hat \p^\dag ]=0\eqno(2.5)$$
Thus we have two sets of independent raising and lowering operators.
We choose to expand $\ve \y \ke$  in the $\{ \ve n,\z \ke \}$ basis,
where $\ve n,\z \ke  \equiv  \ve n \ke  \bigotimes  \ve \z \ke $; i.e
$$\hat \p^\dag \hat \p   \ve n \ke =n \ve n \ke\qquad\hbox{and}\qquad
\hat a \ve \z \ke =\z  \ve \z \ke \eqno (2.6)$$
namely the coherent state. Let
$$\ve \y \ke=\sum _{n=0}^\infty \ve \y_n \ke\qquad\hbox{such that}\qquad
\br m,\bar \z ,t\ve \y \ke = \d_{m,n} \hat \y_n(\bar \z,t) \eqno (2.7)$$
Thus, the action is written as
$$-S_0=\sum_{n=0}^\infty \br \y_n \ve \hat p_t +
\w n \ve \y_n \ke \eqno(2.8)$$
We now introduce electromagnetic perturbations through minimal coupling.
The action is $$-S = \br \y\ve \hat p_t + h_0 + h_{int}\ve \y\ke \eqno (2.9)$$
where
$$ h_{int}={1\over{2m}}[-\vec \P.\vec  {\ba}-\vec {\ba} .\vec \P
+{\vec {\ba}}^2]+A_0.$$
Let
$$A\equiv {1\over{\sqrt{2B}}}({\ba}^x+i{\ba}^y)\qquad\hbox{and}\qquad A^\dag
\equiv {1\over{\sqrt{2B}}}({\ba}^x-i{\ba}^y)$$
which are functions of
$$\hat z\equiv \sqrt {B\over 2}(\hat x+i\hat y)= \hat a-i \hat {\p^\dag}
\qquad\hbox{and}\qquad
\hat{\bar z}\equiv \sqrt {B\over 2}(\hat x-i\hat y)=\hat {a^\dag}+i \hat \p$$
So, $A$, $A^\dag $, $z$, $\bar z$ are all dimensionless. We note,
$$[\hat \p ,A(\hat z,\hat{\bar z})]=-i\del_z A(\hat z,\hat{\bar
z})\qquad\hbox{and}\qquad
[\bar A (\hat z,\hat{\bar z}),\hat\p^\dag]=i\del_{\bar z }\bar A(\hat z,
\hat{\bar z})$$
So, the interaction hamiltonian is rewritten as
$$h_{\rm int}=--{\w \over2}[A\hat \p +\hat{\p^\dag}\bar A+\hat \p A
+\bar A \hat {\p^\dag}] +\w  A{\bar A}+A_0 \eqno(2.10).$$
Now, an arbitrary function $f$ of $\hat z,\ \hat{\bar z}$ may
be written as
$$f(\hat z, \hat{\bar z})=\sum_{n,m}{1\over {{m!}{n!}}}(-i)^n (i)^m
(\hat\p^\dag)^n(\hat\p)^m
\ddagger {\del_z}^n {\del_{\bar z}}^m f(z,\bar z)
\vert_{z={\hat a}^\dag ,\bar z=\hat a}\ddagger$$
where $\ddagger$ $\ddagger$ means antinormal ordering of $\hat a$ and
${\hat a}^\dag $. That is, $\hat a$ is always placed to the left of
${\hat a}^\dag .$
This ordering arises naturally from the normal ordering adopted for $\hat \p$
and ${\hat \p}^\dag .$
Utilizing this, we write
$$ h_{\rm int}=m_{00}+ m_{10}{\hat \p}^\dag + m_{01}{\hat \p}+ m_{11}
{\hat \p}^\dag \hat \p + m_{20} {\hat \p^\dag}{}^2 +m_{02}{\hat \p}^2 +\dots
 \eqno (2.11)$$
$$\eqalignno{m_{00}& ={{i \w }\over2}(\del_z A-\del_{\bar z}\bar A)+\w
A \bar A +A_0+\dots \cr
m_{10}&=-\w \bar A+{\w\over2}\del_z (\del_z A-\del_{\bar z}\bar A)
-i\w \del_z(A \bar A)-i\del_z A_0+ \dots \cr
m_{01}&=-\w A+{\w\over2}\del_{\bar z }(\del_z A-\del_{\bar z}\bar A)
+i\w \del_{\bar z}(A \bar A)+i\del_{\bar z} A_0+ \dots \cr
m_{11}&=i\w (\del_z A-\del_{\bar z}\bar A)+\dots \cr
m_{20}&=i\w \del_z \bar A +\dots \cr
m_{02}&=-i\w \del_{\bar z}A +\dots & (2.12)\cr}$$
The antinormal ordering of $\hat a$ and $\hat {a^\dag}$ is tacit in the
above.
 From (2.9),
$$\eqalign{-S&=\br \y_0 \ve \hat p_t \ve \y_0 \ke +\br \y_0 \ve h_{\rm int}\ve
\y_0 \ke \cr
&+\sum_{n\neq 0}[\br \y_0 \ve h_{\rm int}\ve \y_n \ke +\br \y_n \ve h_{\rm
int}\ve \y_0 \ke] +\sum_{n,n^\prime \neq 0}\br \y_n \ve \hat p_t+
h_{\rm int}+h_0 \ve \y_{n^\prime}\ke \cr}$$
Integrating the modes with $n \neq 0$ out, we get
$$-S_{\rm eff}=\br \y_0 \ve \hat p_t \ve \y_0 \ke +\br \y_0 \ve h_{\rm int}\ve
\y_0 \ke -\br \y_0 \ve h_{\rm int}\hbox{``}{1\over{\hat p_t+h_0+h_{\rm int}}}
\hbox{''} h_{\rm int}\ve \y_0 \ke \eqno(2.13)$$
``${1\over{\hat p_t+h_0+h_{\rm int}}}$'' is just a notation.
It means that
all the intermediate states exclude $n=0$.
So,$$-S_{\rm eff}=\br \y_0 \ve \hat p_t \ve \y_0 \ke +{H_{\rm eff}}^{(0)}
+{H_{\rm eff}}^{(1)}+{H_{\rm eff}}^{(2)}+\br \y \ve h_{\rm int}{1\over {h_0}}
\hat p_t {1\over {h_0}} h_{\rm int}\ve \y \ke +\dots \eqno(2.14).$$
In this expansion and in all subsequent discussions, the occurrence of ${1\over
{h_0}}$ is automatically taken to signify that $n=0$ is omitted
from the intermediate states.
We have dropped the subscript ``0'' from $\ve \y \ke .$
$$\eqalignno{H_{\rm eff}^{(0)}&\equiv \br \y \ve h_{\rm int}\ve \y \ke  \cr
H_{\rm eff}^{(1)}&\equiv -\br \y \ve h_{\rm int}{1\over {h_0}}h_{\rm int}
\ve \y \ke \cr
H_{\rm eff}^{(2)}&\equiv \br \y \ve h_{\rm int}{1\over {h_0}}h_{\rm int}
{1\over {h_0}}h_{\rm int}\ve \y \ke &(2.15) \cr}$$
We choose the external perturbing fields to be slowly varying in space and
time. Thus, we drop all derivatives higher than the leading order. This
means effectively that we may truncate the sum over all $n>0$ to a sum over
the first few terms. Further, since $\ve{{\ba^\m}\over{\sqrt B}}\ve\ll1$, we
drop terms arising out of the higher iterations of
``${1\over{\hat p_t+h_0+h_{\rm int}}}.$''
In computing $H_{\rm eff}$ we come across expressions like
$$\br \y \ve \ddagger A (\hat a, \hat {a^\dag})\ddagger  \ve \y \ke $$
$$\br \y \ve \ddagger A (\hat a, \hat {a^\dag})\ddagger\ \ddagger B(\hat a,
\hat{a}^\dag)\ddagger  \ve \y \ke $$
$$\br \y \ve \ddagger A (\hat a, \hat {a^\dag})\ddagger\ \ddagger B(\hat a,
\hat{a}^\dag)\ddagger\ \ddagger C(\hat a, \hat{a}^\dag)\ddagger \ve \y \ke .$$
These are written as
$$\br\y\ve\ddagger A(\hat a,{\hat a}^\dag)\ddagger\ve\y\ke
=\int d^2 z\ e^{-{\ve z\ve}^2} \bar\y(z,t)\ A(z,\bar z,t)\y (\bar z,t)$$
$$\eqalign{\br\y\ve\ddagger A(\hat a,{\hat a}^\dag)\ddagger\
\ddagger B(\hat a, {\hat a}^\dag)\ddagger\ve\y\ke &=\int d^2 z\
e^{-{\ve z\ve}^2} \bar\y(z,t)\biggl[A(z,\bar z,t)B(z,\bar z,t)\cr
&-\del_{\bar z}A(z,\bar z,t)\del_z B(z,\bar z,t)\dots\biggr]\y (\bar z,t)
\cr}$$
$$\eqalign{\br\y\ve\ddagger A(\hat a,{\hat a}^\dag)\ddagger\
\ddagger B(\hat a, {\hat a}^\dag)\ddagger\
\ddagger C(\hat a, {\hat a}^\dag)\ddagger \ve\y\ke &=\int d^2 z\
e^{-{\ve z  \ve}^2}\bar\y(z,t)\biggl[ABC-A(\del_{\bar z}B)(\del_z C)\cr
&-(\del_{\bar z}A)\del_z (BC)+\dots\biggr]\y(\bar z,t)\cr}$$
where $d^2z\equiv{{d({\rm Re}z)d{(\rm Im}z)}/\p}$.
We write
$$A_0(\vec x,t)=Ex+{\ba}_0(\vec x,t)\eqno(2.16)$$
where $Ex$ is a fixed background.
Using (2.11) to (2.16), it is straightforwardly shown that
$$\eqalignno{-L_{\rm eff}&={1\over2}\int d^2 z\ e^{-{\ve z  \ve}^2}
\biggl[1+A\bar A -{1\over2}(\del_{\bar z}A)(\del_z \bar A)
+{{iE}\over{\w\sqrt{2B}}}(A-\bar A)\biggr]
\biggl(\bar\y (z,t)(-i\adel_t)\y (\bar z,t)\biggr)\cr
&+\int d^2 z\ \r (z,\bar z,t)\biggr[{\ba}_0+{{i\w}\over 2}
(\del_z A-\del_{\bar z}\bar A)-i(A\del_z {\ba}_0-\bar A
\del_{\bar z}{\ba}_0)-iA\del_t \bar A \cr
&+{E\over{\sqrt{2B}}}\biggl((z+\bar z)-
i(A-\bar A)+{1\over2}(z+\bar z)(\del_z A-\del_{\bar z}\bar A)^2+
{1\over2}\del_z\bigl[-(z+\bar z)A(\del_z A-\del_{\bar z}\bar A)-
A^2\bigr]\cr
&+{1\over2}\del_{\bar z}\bigl[(z+\bar z)\bar A(\del_z A-
\del_{\bar z}\bar A)-{\bar A}^2\bigr]+(z+\bar z)A\bar A-{1\over2}
(z+\bar z)(\del_{\bar z}A)(\del_z\bar A)\biggr)\biggr]&(2.17)\cr}$$
where
$$\r(z,\bar z,t)\equiv\bar\y(z,t)\y(\bar z,t)e^{-\ve z\ve^2}
=\br\y\ve z\ke\br z\ve\y\ke e^{-\ve z\ve^2}$$
In deriving (2.17) we have kept upto the quadratic in the elctromagnetic
potentials. Further, we have dropped all spatial and time derivatives of
$ {\ba}^\m $, $A_0$ beyond the appropriate leading order in accordance with our
assumption about the slowly varying nature of $ {\ba}^\m $, $A_0$.
This effective Lagrangian is not yet quite a canonical fermionic Lagrangian.
To get a standard fermionic Lagrangian, we redefine $\y (\bar z,t)$
and $\bar \y (z,t)$ as
$$\y (\bar z,t) \to \ddagger \biggl[1+{1\over2}A\bar A (\del_{\bar z},\bar z,t)
-{1\over4}F(\del_{\bar z},\bar z,t)+{{iE}\over{2\w\sqrt{2B}}}(A-\bar A)
(\del_{\bar z},\bar z,t)\biggr] \ddagger\ \y(\bar z,t)$$
$$\bar \y (z,t) \to \ddagger \biggl[1+{1\over2}A\bar A
(\del_z ,z,t)-{1\over4}F(\del_z,z,t)+{{iE}\over{2\w\sqrt{2B}}}(A-\bar A)
(\del_{z},z,t)\biggr] \ddagger\ \bar \y(z,t) \eqno(2.18)$$
where $F \equiv (\del_{\bar z}A)(\del_z \bar A)$
and $\ddagger \ \ddagger $ indicates that in the first expression in (2.18)
$\del_{\bar z}$ is always to the left of $\bar z$ and in the second
expression $\del_{z}$ is always to the left of $z$.
With these redefinitions  , we get
$$\eqalignno{-L_{\rm eff}&=\int d^2 z\ e^{-{\ve z \ve}^2} \bar
\y(z,t)(-i\del_t)\y (\bar z,t)\cr
&+\int d^2 z\ \r (z,\bar z,t)\biggr[{\ba}_0+{{i\w}\over 2}
(\del_z A-\del_{\bar z}\bar A)-i(A\del_z {\ba}_0-\bar A
\del_{\bar z}{\ba}_0)-{i\over2}(A\del_t \bar A -\del_tA\bar A)\cr
&+{E\over{\sqrt{2B}}}\biggl((z+\bar z)-
i(A-\bar A)+{1\over2}(z+\bar z)(\del_z A-\del_{\bar z}\bar A)^2+
{1\over2}\del_z\bigl[-(z+\bar z)A(\del_z A-\del_{\bar z}\bar A)-
A^2+A\bar A\bigr]\cr
&+{1\over2}\del_{\bar z}\bigl[(z+\bar z)\bar A(\del_z A-
\del_{\bar z}\bar A)-{\bar A}^2+A\bar A\bigr]\biggr)\biggr]&(2.19)\cr}$$

The microscopic action had a local $U(1)$ invariance. The effect of
${\ba}^\m \to {\ba}^\m +\del^\m\Lambda$ was compensated for by suitable
local redefinitions of the phase of the fermionic field.
But within the framework of the effective action, where the fermionic modes for
$n\neq0$ are unavailable, the transformation required for
$\ve\y\ke$ is necessarily more complicated than just a local redefinition
of phase. The transfomation required can be shown to be
$$\eqalignno{\d\y(\bar z,t)&=\ddagger\biggl[-i\L+{1\over2}(\del_{\bar z}\L)
\bar A-{1\over2}A(\del_z\L)-{1\over4}(\del_{\bar z}^2\L)(\del_z\bar A)\cr
&+{1\over4}(\del_{\bar z}A)(\del_z^2\L)+{{iE}\over{2\w\sqrt{2B}}}
(\del_{\bar z}\L+\del_{z}\L)\biggr](\del_{\bar z},\bar z,t)\ddagger\y(\bar
z,t)&(2.20)\cr}$$
Similarly,
$$\eqalignno{\d\bar\y(z,t)&=\ddagger\biggl[i\L-{1\over2}(\del_{\bar z}\L)
\bar A+{1\over2}A(\del_z\L)+{1\over4}(\del_{\bar z}^2\L)(\del_z\bar A)\cr
&-{1\over4}(\del_{\bar z}A)(\del_z^2\L)-{{iE}\over{2\w\sqrt{2B}}}
(\del_{\bar
z}\L+\del_{z}\L)\biggr](\del_{z},z,t)\ddagger\bar\y(z,t)&(2.21)\cr}$$
These are $W_\infty$ transformations [5].
One can show in a straightforward fashion that for infinitesimal $\Lambda$,
which we have assumed all along, (2.20) and (2.21) cancel out all
the $\Lambda$-dependence arising from $S_{\rm eff}$ as ${\ba}^\m\to {\ba}^\m
+\del^\m\Lambda$. Here $\ddagger\ \ddagger$ indicates that in (2.20),
$\del_{\bar z}$ is always to the left of $\bar z$ and in (2.21),
$\del_z$ is always to the left of $z$. Thus $L_{\rm eff}$ in (2.19) is gauge
invariant.
\vfill
\bigskip
\centerline{\bf III. Constructing the droplet.}
\bigskip
If we imagine, momentarily that the
fluctuating potentials have been switched off,
the effective Lagrangian from (2.21) yields an effective action
$$S_{\rm eff}^{(0)}=\br\y\ve \bigl(i\del_t-{E\over\sqrt{2B}}
(\hat a+{\hat a}^\dag)\bigr)\ve\y\ke\eqno(3.1)$$
where $\ve\y\ke$ is a second quantized operator.
But $(\hat a+{\hat a}^\dag)/\sqrt{2B}=\hat X$. This implies that the dominant
part of the effective Hamiltonian is
$$E\br\y\ve\hat X\ve\y\ke
=E\int_{-\infty}^{\infty}dX\ \y^\dag(X)X\y(X)\eqno (3.2)$$
where $\y(X)=\br X\ve\y\ke$ and $\{\ve X\ke\}$ is the basis of the
coordinate representation.
We define the droplet to be such that all the single particle states upto
the zero energy state are filled. So $X=0$ is the Fermi surface. But, for a
large magnetic field, $X\simeq x$, the real spatial coordinate. So, in
physical space, the edge of the droplet is at $x=0$. The ground state as
defined above is
$$\ve G\ke\equiv\prod_{X\leq0}\y^\dag(X)\ve0\ke\eqno(3.3)$$
where $\ve0\ke$ is the Fock vacuum. With respect to the ground state
we re-define the fermion operators appropriately as particle and
antiparticle (hole) operators.

Excitation of this droplet by means of the fluctuating potentials
means the destruction of an electron
within the Fermi sea and the creation of an electron outside of the Fermi
sea. In terms of the state $\ve G\ke$, this translates into the creation of
a neutral particle-antiparticle excitation from the ground state.

Given that the perturbing potential is small and slowly varying in
space-time, (this is the justification for the derivative expansion we have
performed) we would expect only those electrons within some distance
$\l\ll 1/\sqrt B$ to participate in the neutral excitations. In fact,
an expansion in $X$ about $X=0$ yields, to leading order, an action
for fermions interacting with $A$, $\bar A$, ${\ba}_0$ on the boundary of the
droplet $(X=0)$.
The neutral particle-antiparticle excitations mentioned earlier are
actually the neutral excitations around the filled Fermi sea for this
boundary action.

To extract this boundary action from $S_{\rm eff}$, we write the density
operator as
$$\eqalignno{\hat \r(z,\bar z,t)&
=\int_{-\infty}^\infty dX\int_{-\infty}^\infty
dX^\prime\ \br\y\ve X,t\ke\br X^\prime,t\ve\y\ke\br X\ve z\ke\br \bar z\ve
X^\prime\ke e^{-\ve z\ve^2}\cr
&=\int_{-\infty}^\infty dX\int_{-\infty}^\infty dX^\prime\
\y^\dag(X,t)\y(X^\prime,t)\br X\ve z\ke\br\bar z\ve
X^\prime\ke e^{-\ve z\ve^2}&(3.4)\cr}$$
where $\{\y^\dag(X,t),\y(X^\prime,t)\}=\d(X-X^\prime)$.
We note that if $X,X^\prime\leq 0$, $\y^\dag,\y$ exchange their roles with
respect to $\ve G\ke$. This implies that
$$\r(z,\bar z,t)=\int_{-\infty}^\infty dX\int_{-\infty}^\infty
dX^\prime\ [:\y^\dag(X,t)\y(X^\prime,t):+
\Theta(-X)\d(X-X^\prime)]\br X\ve z\ke\br\bar z\ve X\ke
e^{-\ve z\ve^2}\eqno(3.5)$$
where : : indicates normal ordering with respect to $\ve G\ke$. The term
independent of the fermion fields is
$$\int_{-\infty}^0dX\ \br X\ve z\ke\br\bar z\ve X\ke e^{-\ve z\ve^2}=
{1\over{\sqrt\p}}
\int_{-\infty}^{-\sqrt Bx}d\g\ \exp(-\g^2)
\sim\Theta(-x)\eqno(3.6)$$
The last expression is valid in the large B limit.
When this is inserted in (2.19),
we get an effective action involving only $\ba_\m$, called the bulk action.
$$\eqalignno{S_{\rm eff}^{\rm bulk}&\simeq
-{B\over{2\p}}\int_{-\infty}^{\infty}dt
\int_{-\infty}^{0}dx\ \int_{-\infty}^{\infty}dy\
\biggl[Ex+ {\ba}_0-{1\over{2m}}{\bb}+{E\over B}{\ba}^y
+{1\over {2B}}\e^{\m\n\r}{\ba}_\m\del_\n{\ba}_\r&(3.7)\cr
&-{{2E}\over{(2B)^2}}\biggl\{2{\ba}^x\del_y {\ba}^y+x( {\ba}^x\del_y
{\bb}-{\ba}^y\del_x{\bb})-3{\ba}^y{\bb}\biggr\}\biggr]
+{1\over{4\p}}\int_{-\infty}^{\infty}dt
\int_{-\infty}^{\infty}dy\ {\ba}^0(y,t){\ba}^y(y,t)\cr}$$
where ${\bb}\equiv\del_x{\ba}^y-\del_y{\ba}^x,$ ${\ba}^\m (y,t) \equiv {\ba}^\m
(x=0,y,t).$
We see that $S_{\rm eff}^{\rm bulk}$ contains a Chern-Simons term.
It exists only in the bulk of the droplet.
Under gauge transformation, ${\ba}^i\to {\ba}^i-\del_i\Lambda$ and
${\ba}^0\to {\ba}^0+\del_t\Lambda$,
$$\d S_{\rm eff}^{\rm bulk}=-{1\over {2\p}}\int_{-\infty}^{\infty}dt
\int_{-\infty}^{\infty}dy\ \Lambda(y,t)\bigl[\del_y{\ba}_0(y,t)+
{E\over B}\del_y{\ba}^y(y,t)\bigr]\eqno (3.8)$$
So the gauge dependence is through a boundary term.
The fermionic part of $S_{\rm eff}$ is obtained by inserting
$\colon \r(z, \bar z , t) \colon $ in place of $\r(z, \bar z , t) $
in (2.19). We expand the fermionic part of $S_{\rm eff}$ and keep only
the low momentum $(X\sim0)$ fermions. Thus we get
$$S_{\rm bdry} = \int_{-\infty}^{\infty} dt\int_{-\infty}^{\infty}
dY\ \y^\dag (Y,t) \biggl[ \{ i \del_t - {\ba}_{0}(Y,t) \} +
{E\over B} \{-i \del_{Y}-{\ba}^y(Y,t)\}\biggr] {\y} (Y,t).
\eqno(3.9)$$
Here normal ordering with respect to the ground state is implicit.
$\y(Y,t)\equiv\br Y\ve\y\ke$ is actually the Fourier transform of
$\br X\ve\y\ke$ used in (3.4), since $[\hat X,\hat Y]={i\over B}$. We
however continue to use the same symbol $\y$. Thus
$$\int_{-\infty}^{\infty} dt\ \int {{d^2 \z}} e^{-{\ve \z \ve}^2}
\y^\dag (\z, t) i\del_t {\y} (\bar \z, t) =
\int_{-\infty}^{\infty} dt\ \int_{-\infty}^{\infty} dY\ \y^\dag (Y, t)
i\del_t {\y} (Y, t).\eqno(3.10)$$
Also,
$$\br \y \ve E \hat X \ve \y \ke = i {E\over B}
\y^\dag (Y, t) \del_{Y} {\y} (Y, t).\eqno(3.11)$$
which we have used in deriving (3.9).
The neutral excitations of electron--hole pairs are the neutral
excitations of this boundary action since we have expressed the low
momentum $(X \simeq 0)$ part of the original normal ordered fermion
action as this boundary action.  The electrons in the high momentum eigen
states are unaffected as long as the momenta of the
perturbing potentials are $\ll \sqrt B$.
Since we are interested in the low energy perturbations of the droplet,
$S_{\rm eff}^{\rm bulk}$ is the net effect of the fermions inside of the
droplet. $S_{\rm bdry}$ is classically gauge invariant. But quantum
mechanically this theory is an anomalous gauge theory [7,8].
 This means that the quantized theory will not be gauge invariant. The gauge
parameter dependence of the theory can be best extracted by writing it in the
bosonic language.

By studying the bosonization of the L.L.L. fermions, the following result was
obtained in [5]. The Lagrangian is written in the form of (2.19) in terms of
the L.L.L. fermion fields, namely
$$L_{\rm eff}= \int d\vec x\  \bar \y(\vec x,t)i\del_t \y(\vec x,t) -\int d\vec
x\  \bar \y(\vec x,t) \y(\vec x,t) f(\vec x,t)\eqno(3.12)$$
where $f$ is a function of $\vec{\ba}$ and ${\ba}_0$, as given in
(2.19) and $\y(\vec x,t)$ satisfies the L.L.L. condition, $(\del_z
+{1\over2}\bar z)\y(\vec x,t)=0$.
In the bosonized form,these fields are replaced by bosonic Schr\"odinger
fields.
In the bosonized form, the Lagrangian is precisely given by (3.12), where $\y$
is
replaced by a bosonic Schr\"odinger wave field which obeys the nonlinear L.L.L.
condition,
$ \bigl(\del_z +{1\over2}\bar z -\int {{{\bar \y}\y({\vec
x}^\prime)}\over{z-z^\prime}}\bigr)\y(\vec x,t)=0$
. This L.L.L. condition can be approximately solved in the droplet
approximation. This solution is that the density ${\bar \y}\y(\vec x)$ is equal
to $B/2\p$ inside of a certain region of space (droplet) and
zero outside.
The dynamical variable is then the boundary fluctuations
of the uniform density around this ``classical'' droplet
configuration. We denote these fluctuations in the following by $\f(Y,t)$.
The first term in (3.12) gives the first term in the free chiral bosonic
Lagrangian [5],[9],[10],
in terms of $\f(Y,t)$.
Working within this droplet approximation, we obtain
$$\eqalignno{L_{\rm eff}&={{B^2}\over {8\p}}\int dYdY^\prime\ \f (Y,t)\
\e (Y-Y^\prime)\dot\f(Y^\prime,t)-{B\over {2\p}}\int dY\bigl[-{\ba}_0
-{E\over B}{\ba}^y\bigr]\f(Y)\cr
&-{{BE}\over{2\p}}\int dY\ {1\over2}\f^2(Y)-{B\over{2\p}}\int_{-\infty}^0dx
\int_{-\infty}^{\infty}dy\ \biggl[{\ba}_0-{1\over{2m}}{\bb}+Ex+{E\over B}
{\ba}^y\cr
&+{1\over {2B}}\e^{\m\n\r}{\ba}_\m\del_\n{\ba}_\r
-{{2E}\over{(2B)^2}}\biggl\{2{\ba}^x\del_y{\ba}^y+
x({\ba}^x\del_y{\bb}-{\ba}^y\del_x{\bb})-3{\ba}^y{\bb}\biggr\}
\biggr]\cr
&+{1\over{4\p}}\int_{-\infty}^{\infty}dt
\int_{-\infty}^{\infty}dy\ {\ba}^0(y,t){\ba}^y(y,t) &(3.13)\cr &}$$
where $\e(Y-Y^\prime)$ is equal to 1 for $Y>Y^\prime$ and $-1$
for $Y<Y^\prime$.
The equation of motion for $\phi(Y,t)$ is
$${{B^2}\over{4\p}}\int dY^\prime\ \e(Y-Y^\prime)\ \dot\phi (Y^\prime,t)
-{{BE}\over {2\p}}\f(Y,t)+{B\over{2\p}}({\ba}_0+{E\over B}{\ba}^y)=0
\eqno(3.14)$$
This means quantum mechanically
$$\bigl[\del_t-{E\over B}\del_Y\bigr]\bigl\langle\f(Y,t)\bigr\rangle
=-{1\over B}
\del_Y\bigr({\ba}_0+{E\over B}{\ba}^y\bigr)\eqno(3.15)$$
where $\bigl\langle\dots\bigr\rangle$ denotes the quantum mechanical
average over $\f$.
Now under gauge transformation, the change in the partition function
due to the change in the action for $\f$ is
$$\eqalignno{\bigl\langle\d S_{\rm eff}^{\rm bdry}\bigr\rangle&
-{B\over{2\p}}\int dt\int dY\ \bigl[
-\del_t\Lambda+{E\over B}\del_Y\Lambda\bigr]\bigl\langle\f(Y,t)\bigr\rangle
\cr
&=-{B\over{2\p}}\int_{-\infty}^\infty dt \int_{-\infty}^\infty dY\
\Lambda (Y,t)\bigl(\del_t-{E\over B}\del_Y\bigr)\bigl\langle\f(Y,t)\bigr
\rangle\cr
&={1\over{2\p}}\int_{-\infty}^\infty dt \int_{-\infty}^\infty dY\
\Lambda(Y,t)\bigr[\del_Y{\ba}_0+{E\over B}\del_Y{\ba}^y\bigr]&(3.16)\cr}$$
by (3.15).
So comparing (3.8) and (3.16) we see that all the
$\Lambda$-dependence precisely cancels out.
We have therefore demonstrated explicitly that all the gauge non-invariance
of the bulk, which appears as a boundary term is precisely removed by the
gauge non-invariance of the chiral bosonic action governing the surface
oscillations of the droplet.

This emphasizes the importance of the edge states in maintaining gauge
invariance in the system.
\vfill
\bigskip
\centerline{\bf V.  Conclusion}
\bigskip
In this paper we have studied the electromagnetic interactions
of a quantum Hall droplet. We integrate the higher Landau levels out
and within
the framework of a derivative-expansion scheme obtain a gauge invariant
effective action
for the electrons in the L.L.L.  We know that $2+1$ dimensional electrons in
the L.L.L.
are equivalent to $1+1$ dimensional electrons, the configuration space of
the $2+1$ dimensional system being the phase space of the $1+1$ dimensional
system, [5]. In the Thomas-Fermi picture [11], the $1+1$ dimensional electron
gas occupies a region of constant density in phase space. This region in the
$1+1$ deimensional phase space coincides precisely with the physical droplet
of electrons in the L.L.L. created by the background electrostatic
potential.
The bulk of the droplet which
corresponds to a filled fermi sea contributes an effective action, called the
bulk action, in terms
of the perturbative electromagnetic potentials.
The bulk action is not gauge invariant. This
non-invariance, however, is spurious as it is cancelled by the gauge
non-invariance
of the $1+1$ dimensional edge system. Thus the basic
mechanism for the preservation of gauge invariance as suggested in [2], [3]
and [4] is seen to be valid. Moreover, we have a well-defined and systematic
procedure for isolating the $1+1$ dimensional edge from the original $2+1$
dimensional system.

The case of the fractional Hall droplet can be similarly handled, at least
phenomenologically, by writing the bosonic density ${\bar \y}\y (\vec x) $ in
the discussion following (3.12), as $\n B/2\p$ inside of the
droplet, where $\n$ is the appropriate filling factor.
\bigskip
\bigskip
\centerline{\bf Acknowledgements}
\bigskip
We acknowledge useful discussions with S. Iso, D. Karabali,
P. Orland and P. Panigrahi.
This work was supported by the NSF grant PHY90-20495 and the Professional
Staff
Congress Board of Higher Education of the City
University of New York under grant
no. 6-63351.
\vfill\eject
\bigskip
\centerline{\bf References}
\bigskip
\item{[1]}  B.I. Halperin, Phys. Rev. {\bf B 25}, 2185, (1982)
\item{[2]}  X.G. Wen, Phys. Rev. {\bf B 43}, 11025, (1990) and references
therein
\item{[3]}  M. Stone, Ann. Phys. (N.Y.), {\bf 207}, 38, (1991);
Int. J. Mod. Phys. {\bf B5}, 509, (1991)
\item{[4]}  J. Fr\"ohlich \& T. Kerler, Nucl. Phys. {\bf B354}, 365, (1991)
\item{[5]}  S. Iso, D. Karabali \& B. Sakita, Nucl. Phys. {\bf B 388},
700, (1992); Phys. Lett.  {\bf B 296}, 143, (1992)
\item{[6]}  S.M. Girvin \& T. Jach, Phys. Rev. {\bf B 29}, 5617, (1983)
\item{[7]}  R. Jackiw \& R. Rajaraman, Phys. Rev. Lett. {\bf 54}, 1219, (1985)
\item{[8]}  L. Faddeev \& S. Shatashvili, Phys. Lett. {\bf 167 B} 225,(1986)
\item{[9]}  R. Floreanini \& R. Jackiw, Phys. Rev. Lett. {\bf 59},1873, (1987)
\item{[10]} J. Sonnenschein, Nucl. Phys. {\bf B309}, 752, (1988)
\item{[11]}  See e.g. C. Kittel, ``Introduction to Solid State Physics, ''
(Wiley, New York, 1986)
\end